\documentclass[11pt]{article}
\usepackage{epsfig,dsfont}
\usepackage{geometry}
\usepackage[normalem]{ulem}
\usepackage{color}

\usepackage{cite}

\begin{document}  
\sffamily

\vspace*{1mm}

\begin{center}

{\LARGE
Dual lattice representations for O(N) and CP(N-1) models \\
\vskip2mm
with a chemical potential}
\vskip8mm
Falk Bruckmann$\,^a$, Christof Gattringer$\,^b$, Thomas Kloiber$\,^{a,b}$ and Tin Sulejmanpasic$\,^c$ 
\vskip8mm
$\;^a$ Universit\"at Regensburg, Institut f\"ur Physik, Universit\"atstra{\ss}e 31, 93053 Regensburg, Germany
\vskip1mm
$\;^b$ Universit\"at Graz, Institut f\"ur Physik, Universit\"atsplatz 5, 8010 Graz, Austria
\vskip1mm
$\;^c$ North Carolina State University, Department of Physics, Raleigh, NC 27695-8202, USA
\end{center}
\vskip10mm

\begin{abstract}
We derive dual representations for O(N) and CP(N-1) models on the lattice. In terms of the dual variables the partition sums
have only real and positive contributions also at finite chemical potential. Thus the complex action problem of the conventional formulation
is overcome and using the dual variables Monte Carlo simulations are possible at arbitrary chemical potential.
\end{abstract}
\vskip5mm
\begin{center}
{\sl Journal Reference: Phys.~Lett.~B749 (2015) 495-501}
\end{center}

\vskip10mm

\section{Introductory remarks}
A very visible shortcoming of lattice QCD is the current inability to properly deal with QCD at finite density. This is not a fundamental 
conceptual problem, but more of a technical challenge: At finite chemical potential $\mu$ the action $S$ is complex and the Euclidean 
Boltzmann factor $e^{-S}$ does not have a probability interpretation which is necessary for a Monte Carlo simulation. As a matter of fact, 
the complex action problem appears in most field theories at finite chemical potential, both on the lattice and in the continuum, and for 
fermionic as well as for bosonic degrees of freedom. In these systems importance sampling methods are thus not able to identify `relevant
configurations' among the fundamental degrees of freedom, which might hint at more effective degrees of freedom allowing
for a better understanding of physical effects at finite density. 
As a side remark we also note that the addition of a topological term leads
to a complex action problem, similar to the one at finite $\mu$.

A very powerful approach to completely solving the complex action problem is to exactly rewrite the partition sum in terms of new 
degrees of freedom such that the partition sum has only real and positive contributions. These new degrees of freedom are often referred to
as dual variables which are given by loops for matter fields and by surfaces for gauge fields. In recent years several interesting lattice
field theories at finite chemical potential or with a topological term were rewritten in terms of dual variables 
(for two reviews see, e.g., \cite{review1,review2}). In the dual formulation the system can, e.g., be updated efficiently with worm algorithms of the 
Prokofev-Svistunov type \cite{worm}, which also has a generalization to the surfaces that dualize the gauge fields \cite{surfaceworm}.

An important lesson that has been learned when developing the dual approach is that there is no general strategy for finding a 
dual representation. Every type of model has to be analyzed individually and in particular the type of symmetry plays an important role
since it directly determines the constraints for the dual variables which in turn determine the structure of the dual loops and surfaces. A
second insight that has emerged is that a dual representation is not unique -- there might be several different dual representations for the same
lattice field theory.  The different dual representations may have different properties, in particular with respect to a possible 
dual Monte Carlo simulation. 

An important class of systems are lattice field theories with O($N$) symmetric actions and the closely related CP($N\!-\!1$) models. 
These models are especially interesting in two dimensions because of their close resemblance to gauge theories. In particular both 
O($N$) and CP($N\!-\!1$) in two (1+1) dimensions are asymptotically free and have a dynamically generated mass gap proportional to 
the strong scale of the theory. In addition O($3$) and CP($N\!-\!1$) for any $N$ have topological charge and instantons and allow 
for a $\theta$ term, making them even more attractive for understanding the gauge dynamics. 

We mention in passing that various chemical potentials which we discuss here have a close similarity 
with twisted boundary conditions for the fields of the model (the chemical potentials  can 
be analytically continued to the twists), and were studied extensively in the context of fractional topological charge and resurgence
\cite{Eto:2004rz,Eto:2006mz,Bruckmann:2007zh,Brendel:2009mp,Harland:2009mf,Dunne:2012ae,Dunne:2012zk,Cherman:2013yfa,Nitta:2014vpa,Nitta:2015tua,Dunne:2015ywa}.

For models with these symmetries, dual variants were discussed in the literature 
\cite{review1,endres,wolff_ON,wolff_CPN,shailesh_O2,alles,falktin,meurice}. 
Here we revisit the problem of dualizing O($N$) and CP($N\!-\!1$) models with finite chemical potential. We show that both systems allow for a 
dualization along the same lines using $N$ sets of dual flux variables for the O($N$) case and $2N$ sets for CP($N\!-\!1$). 
It is interesting to note that the dualizations presented here differ fundamentally from other suggestions (which only considered the case 
without chemical potential): For the O($N$) model a dualization with only one set of dual variables was presented in \cite{wolff_ON} (we use 
$N$ sets of dual variables), while for the CP($N\!-\!1$) case variants with 2 \cite{wolff_CPN}  and with $N^2$ \cite{review1} sets of dual variables 
were discussed (we use $2N$ sets). 

The fact that rather different exact dual representations are available for the same model is interesting for at least two reasons: Identifying
a real and positive dual representation is only the first step of the dual approach. One also has to find a suitable Monte Carlo algorithm for
updating the dual degrees of freedom which are subject to constraints that have to be obeyed during the update.  
Different representations that use completely different dual variables will necessarily give rise to Monte Carlo update schemes with different
properties and different efficiency. Exploring the space of possible dual representations is certainly important for identifying the most efficient 
schemes. The second reason for exploring different dual representations is that currently for the probably most interesting case 
of non-abelian gauge systems so far only dual representations that also have negative weights were found
\cite{dualsun_1,dualsun_2,dualsun_3,dualsun_4,dualsun_5,dualsun_6,dualsun_7,dualsun_8}. However, with every new dualization strategy we 
enlarge our toolbox and get closer to the goal of finding real and positive representations suitable for dual Monte Carlo simulations.

\section{Lattice formulation of O($N$) and CP($N\!-\!1$) models with a chemical potential}

In this section we first discuss the conventional lattice discretization for the O($N$) model with chemical potential and then 
for the CP($N\!-\!1$) model. For vanishing chemical potential the action of the nonlinear O($N$) model on the lattice is given by  \begin{equation}
S[\vec{r}\,] \; = \; -J \sum_{x\in\Lambda} \sum_{\nu=1}^{d} \vec{r}\,(x) \! \cdot \vec{r}\, (x+\hat{\nu}) \; ,
\label{action0}
\end{equation}
where the first sum runs over all sites of the $d$-dimensional lattice $\Lambda$ with size $V = N_s^{d-1} \times N_t$ 
and periodic boundary conditions in all directions.
The second sum runs from $\nu = 1$ to $\nu = d$, and by $\hat{\nu}$ we denote the unit vector in $\nu$-direction. The direction 
$\nu = d$ will be referred to as the Euclidean time direction.
The dynamical degrees of freedom are real $N$-component spin vectors $\vec{r}\,(x)$ assigned to the sites of the lattice. The
vectors obey the constraints $\vec{r} \, (x)^2 = 1$, i.e., they are constrained to the \(N-1\) sphere $S^{N-1}$. The action (\ref{action0}) 
consists of a ferromagnetic nearest neighbor interaction of the spin vectors with a coupling strength $J$ (the lattice spacing $a$ is set 
to $a=1$ throughout this paper). The partition sum is given by $Z=\int \mathcal{D}[\vec{r}\,] \exp(-S[\vec{r}\,])$ and the coupling $J$ is positive 
(ferromagnetic case). The integration measure $\mathcal{D}[\vec{r}\,]$ is the product of the O($N$) invariant measures 
over $S^{N-1}$ at each site (see (\ref{on_measure}) below).

Due to the global O($N$) symmetry the model has conserved charges which can be coupled to chemical potentials. 
Following \cite{Hasenfratz:1990ab,falktin} we introduce a chemical potential $\mu$ to one of them (all other cases can be obtained by adding 
several such ''fundamental" chemical potentials \cite{falktin}). The chemical potential gives a different weight to forward
and backward nearest neighbor terms in time direction ($\nu = d$). The action with chemical potential thus reads 
\begin{eqnarray}
\label{eq_action_mu}
S[\vec{r}\,] &=& -J\sum_{x\in\Lambda} \sum_{\nu=1}^{d} \Big[ \vec{r}_{\perp}(x)\cdot \vec{r}_{\perp}(x+\hat{\nu})\nonumber \\
&+ & \frac{1}{2} \,  r_{12}(x) \,  r_{12}(x+\hat{\nu}) \left(e^{-\mu\delta_{\nu,d}} e^{-i(\phi(x)-\phi(x+\hat{\nu}))} + 
e^{\mu\delta_{\nu,d}} e^{i(\phi(x)-\phi(x+\hat{\nu}))}\right)\Big] \; .
\end{eqnarray}
We use a notation where the first two components are combined into a complex number $r_1(x)+i r_2(x)=r_{12}(x) \, e^{i \phi(x)}$ 
and $\vec{r}_{\perp}(x)$ 
contains the $N-2$ components which do not couple to the chemical potential,
\begin{equation}
\vec{r}_\perp \; = \; (r_3(x),~r_4(x),\ldots r_N(x)) \; .
\end{equation}
We require $r_{12}(x)^2 + \vec{r}_{\perp}(x)^2 = 1$ and thus for $\mu = 0$ the action (\ref{eq_action_mu}) reduces to (\ref{action0}). 
We stress that in the conventional form  the action is complex for $\mu \neq 0$ and a Monte Carlo simulation 
suffers from the complex action problem.

\vskip5mm

Let us now come to the CP($N\!-\!1$) model where various different lattice discretizations were discussed (see, e.g., \cite{wolff_CPN,cpnlat}). 
We here use a variant where the interaction is introduced with an auxiliary link field $U_\nu(x)$. 
For CP($N\!-\!1$) the chemical potentials can be added for the charges associated with the global SU($N$) symmetry. 
To access the corresponding phase
diagram experimentally, a set-up relying on atoms in optical lattices has been proposed recently \cite{zoller}. These models enjoy a global U($N$) 
symmetry, but the U(1) part of it is gauged with non-dynamical gauge fields, so that the Gauss constraint is the statement that the charge density, 
and therefore charge, vanishes identically. This means that a chemical potential for U(1) has no effect on the theory, as the U(1) charge is forbidden. 
We will see a manifestation of this in the dual variables as well. Generic chemical potentials can be rotated to those coupling to the $N-1$ diagonal 
generators of the symmetry. The corresponding lattice action 
reads (by  $\mu_j, j = 1,2 \, ... \, N$ we denote the chemical potentials for the $N$ components),
\begin{equation}
S[\vec{z}, U] \; = \; - J \sum_{j=1}^N \sum_{x\in\Lambda} \sum_{\nu=1}^{d} \Big[ e^{- \mu_j \, \delta_{\nu,d}} z_j(x)^* \, U_\nu(x) z_j(x+\hat{\nu}) 
+ e^{\, \mu_j \, \delta_{\nu,d}} z_j(x) \, U_\nu(x)^* z_j(x+\hat{\nu})^* \Big] \; .
\label{eq_actioncpn}
\end{equation}
The degrees of freedom are $N$-component complex vectors with components $z_j(x) \in \mathds{C}$. The spins are normalized to 1, i.e., 
$\sum_{j=1}^N{|z_j(x)|^2} = 1$. As for the O($N$) model we consider a $d$-dimensional $N_s^{d-1} \times N_t$ lattice $\Lambda$ with periodic 
boundary conditions. The nearest neighbor terms of the action are connected with U(1) link variables, $U_\nu(x) = e^{i A_\nu(x)}$ with
$A_\nu(x) \in [0,2\pi]$. In the temporal direction ($\nu = d$) the terms $e^{\, \pm \mu_j \, \delta_{\nu,d}}$ are non-trivial and give a
different weight to forward and backward propagation in time. 

The partition sum is given by $Z=\int \mathcal{D}[U] \mathcal{D}[\vec{z}\,] \exp(-S[\vec{z},U])$ and the coupling $J$ is positive 
(ferromagnetic case). The integration measure $\mathcal{D}[\vec{z}\,]$ is the product over the measures for the $\vec{z}(x)$ at each lattice site
and $\mathcal{D}[U]$ is the product of Haar measures for the $U_\nu(x) \in$ U(1) at all links (see (\ref{cpn_measure}) 
and (\ref{link_integration}) below for the explicit form). 

\section{Dual representation of O($N$) models with a chemical potential}

Both actions (\ref{eq_action_mu}) and (\ref{eq_actioncpn}) obviously have a complex action at non-zero chemical potentials. 
In this and the next section we discuss the exact mapping of the corresponding partition sums to dual variables. We begin with
the O($N$) model. Some of the steps here will then also be  used in the next section for the CP($N\!-\!1$) model.

As the first step towards a dualization we introduce an explicit parametrization of the spin components
$r_{12}(x)e^{i\phi(x)}$ and $\vec{r}_{\perp}(x)$ from (\ref{eq_action_mu}) in terms of generalized spherical coordinates, 
\begin{eqnarray}
r_{12}(x) &=& \sin\theta_N(x) \ldots \sin\theta_4(x) \sin\theta_3(x) \; , \nonumber\\
r_3(x) &=& \sin\theta_N(x) \ldots \sin\theta_4(x)\cos\theta_3(x) \; , \nonumber\\
r_4(x) &=& \sin\theta_N(x) \ldots \sin\theta_4(x)\cos\theta_4(x) \; , \nonumber\\
\vdots && \vdots \nonumber\\
r_{N-1}(x) &=& \sin\theta_N(x) \cos\theta_{N-1}(x) \; , \nonumber\\
r_N(x) &=& \cos\theta_N(x) \; .
\end{eqnarray}
The ranges of values for the angles are \(\phi(x) \in[0,2\pi)\) and \(\theta_i(x) \in [0,\pi]\). The path integral measure 
is a product measure over each individual spin and in terms of the generalized spherical coordinates has the explicit form
\begin{eqnarray}
\label{on_measure}
\int \! \mathcal{D}[\vec{r}\,] &= & \prod_{x} \int_{S^{N-1}} \!\!\!\!\! d^N r(x) \; ,\\
\int_{S^{N-1}} \!\!\!\!\! d^N r(x)&= & \int_0^{2\pi} \! \frac{d\phi(x)}{2\pi} \int_0^\pi \!\! \sin\theta_3(x) ~ 
d\theta_3(x) \int_0^\pi \!\! \sin\theta_4(x)^2 d\theta_4(x) \ldots \int_0^\pi\!\! \sin\theta_N(x)^{N-2} d\theta_N(x) \; .
\nonumber
\end{eqnarray}
We remark that an irrelevant normalization constant was dropped. 

For a convenient notation the action is written as the sum of terms $S^{\,(j)}_{\nu}(x)$ given by
\begin{eqnarray}
S^{\,(1)}_{\nu}(x) & = & \frac{J}{2} \;  r_{12}(x) \, r_{12}(x+\hat{\nu}) \; e^{-\mu\delta_{\nu,0}} \; e^{-i(\phi(x)-\phi(x+\hat{\nu}))} \; , 
\nonumber \\
S^{\,(2)}_{\nu}(x) & = & \frac{J}{2} \; r_{12}(x) \,  r_{12}(x+\hat{\nu}) \; e^{\mu\delta_{\nu,0}} \; e^{i(\phi(x)-\phi(x+\hat{\nu}))} \; , 
\nonumber \\
S^{\,(j)}_{\nu}(x) & = & J \; r_j(x) \, r_j(x+\hat{\nu}) \; , ~~~~~ j=3,4 \, ... \, N \; .  
\end{eqnarray}
The action thus reads $S[\vec{r}\,] = -\sum_{x,\nu,j} S^{\,(j)}_{\nu}(x)$, 
where each term $S^{\,(j)}_{\nu}(x)$ is a simple product of neighboring spin components.

We now factorize the Boltzmann weight and expand each factor individually:  
\begin{eqnarray}
Z & = & 
\int \! \mathcal{D}[\vec{r}\,] \; e^{-S[\vec{r}\,]} \; = \; 
\int \! \mathcal{D}[\vec{r}\,]  \; e^{\; \sum_{x,\nu,j} S^{\,(j)}_\nu(x)} \; = \; 
\int \! \mathcal{D}[\vec{r}\,]  \; \prod_{x,\nu,j} e^{S^{\,(j)}_\nu(x)} 
\nonumber\\
& = & \int \! \mathcal{D}[\vec{r}\,] \prod_{x,\nu}  
\left( \sum_{k_{x,\nu}=0}^\infty \frac{\left(S^{\,(1)}_\nu(x)\right)^{k_{x,\nu}}}{k_{x,\nu}!} \right) 
\left( \sum_{l_{x,\nu}=0}^\infty \frac{\left(S^{\,(2)}_\nu(x)\right)^{l_{x,\nu}}}{l_{x,\nu}!} \right)
\left(\prod_{j=3}^N \sum_{n^{(j)}_{x,\nu}=0}^\infty 
\frac{\left(S^{(j)}_\nu(x)\right)^{n^{(j)}_{x,\nu}}}{n^{(j)}_{x,\nu}!} \right)
\nonumber \\
& = & \sum_{\{n^{(j)},k,l\}} \left(\prod_{x,\nu} 
\frac{(J/2)^{k_{x,\nu}+l_{x,\nu}}}{k_{x,\nu}!~l_{x,\nu}!} \right) 
\left(\prod_{j=3}^{N} \frac{J^{\; n_{x,\nu}^{(j)}}}{n_{x,\nu}^{(j)}!}\right) 
e^{-\mu\sum_x (k_{x,d}-l_{x,d})} ~\mathcal{I}[\{k,l,n^{(j)}\}]~.
\label{expansion1}
\end{eqnarray}
The individual exponentials of the terms $S^{(j)}_{\nu}(x)$ of the action have been expanded with summation variables attached to the
links of the lattice. We use $k_{x,\nu} \in \mathds{N}_0$ and $l_{x,\nu} \in \mathds{N}_0$ for the 
first two terms ($j = 1,2$) and $n_{x,\nu}^{(j)} \in \mathds{N}_0$ for the terms with $j=3,4 \, ... \, N$. In the last line of (\ref{expansion1}) the explicit 
expressions for the $S_\nu^{(j)}(x)$ have been inserted.  After ordering the various factors, the remaining integrals over the angles at the  
individual lattice sites are collected in $\mathcal{I}[\{k,l,n^{(j)}\}]$:
\begin{eqnarray}
&&\;\hspace*{-8mm} \mathcal{I}[\{k,l,n^{(j)}\}] = \prod_x  \!
\int_0^{2\pi} \!\! \frac{d\phi(x)}{2\pi} e^{-i\phi(x) \sum_\nu [k_{x,\nu}-l_{x,\nu}-(k_{x-\hat{\nu},\nu}-l_{x-\hat{\nu},\nu})]}  \!\!
\int_0^\pi \!\!\!\! d\theta_3(x) \! \cos\theta_3(x)^{a_x^{(3)}} \!\! \sin\theta_3(x)^{1+a^{(2)}_x}   \\
&&\;\hspace*{-7mm}  \times  \int_0^\pi \!\!\!\! d\theta_4(x) \cos\theta_4(x)^{a_x^{(4)}} \sin\theta_4(x)^{2+a^{(2)}_x + a_x^{(3)}} 
 \; .... \; 
 \int_0^\pi \!\!\!\! d\theta_N(x) \cos\theta_N(x)^{a_x^{(N)}} \sin\theta_N(x)^{N-2+a^{(2)}_x + a_x^{(3)} \, ... \, a^{(N-1)}_x}\!\!,
 \nonumber
\end{eqnarray}
where we use the following shorthand notation
\begin{equation}
a_x^{(2)} \equiv \sum_\nu \big[ k_{x,\nu}+k_{x-\hat{\nu},\nu}+l_{x,\nu}+l_{x-\hat{\nu},\nu}\big]  \quad , \qquad
a_{x}^{(j)} \equiv \sum_\nu \big[ n_{x,\nu}^{(j)}+n_{x-\hat{\nu},\nu}^{(j)}\big] \; , \;  j = 3,4 \, ... \, N ~.
\end{equation}
The integrals over the azimuthal angles $\phi(x)$ give rise to Kronecker deltas which in turn enforce the constraints
$\sum_\nu [k_{x,\nu}-l_{x,\nu}-(k_{x-\hat{\nu},\nu}-l_{x-\hat{\nu},\nu})] = 0 \; \forall x$.
The integrals over the polar angles are of the form ($a,b \in \mathds{N}_0$)
\begin{equation}
I(a,b) \equiv \int_0^\pi d\theta \big(\cos\theta\big)^a \big(\sin\theta\big)^b \; = \; \left\{
\begin{array}{lc}
\frac{ \Gamma \left( \frac{a+1}{2} \right) \Gamma \left( \frac{b+1}{2} \right) }{ \Gamma \left( \frac{a+b+2}{2} \right)} & \quad a~\mbox{even}\\
0 & \quad a~\mbox{odd}
\label{integrali}
\end{array}
\right. \; ,
\end{equation}
and therefore also impose constraints which restrict the admissible configurations of the variables $n_{x,\nu}^{(j)} \in \mathds{N}_0$. 
The structure of the constraints can be simplified by the following change of variables:
\begin{equation}
k_{x,\nu}-l_{x,\nu} \equiv m_{x,\nu}  \;\;, \;\;\; m_{x,\nu} \in \mathds{Z} \quad \mbox{and} \quad
k_{x,\nu}+l_{x,\nu} \equiv |m_{x,\nu}|+2~\overline{m}_{x,\nu} \;\; , \;\;\;  \overline{m}_{x,\nu} \in \mathds{N}_0 \; .
\end{equation}
The partition function in dual representation assumes the form
\begin{eqnarray}
Z & = & \!\!\! \sum_{\{m,\overline{m},n^{(j)}\!\}}  \left(
\prod_x  \, \delta\Big(\sum_\nu[m_{x,\nu}-m_{x-\hat{\nu},\nu}]\Big) ~ \prod_{j=3}^N E\left(a_x^{(j)}\right) \right)
\label{eq_dual_z} \\
& \times &\quad  
e^{-\mu\sum_x m_{x,d}} 
\left( \prod_{x,\nu} \frac{(J/2)^{|m_{x,\nu}|+2~\overline{m}_{x,\nu}}}{(|m_{x,\nu}|+\overline{m}_{x,\nu})!~\overline{m}_{x,\nu}!} 
\prod_{j=3}^{N} \frac{J^{\; n_{x,\nu}^{(j)}}}{n_{x,\nu}^{(j)}!}\right) 
\left( \prod_x \prod_{j=3}^N I\Big(a_x^{(j)},j-2+\sum_{k=2}^{j-1} a_x^{(k)}\Big) \right),
\nonumber 
\end{eqnarray}
with the $a^{(j)}_x$ now given by 
\begin{equation}
a^{(2)}_x \; = \;  \sum_\nu \big[ |m_{x,\nu}|+|m_{x-\hat{\nu},\nu}|+2(\overline{m}_{x,\nu}+\overline{m}_{x-\hat{\nu},\nu}) \big]  \quad , \qquad
a_{x}^{(j)} \; = \; \sum_\nu \big[ n_{x,\nu}^{(j)}+n_{x-\hat{\nu},\nu}^{(j)} \big] \; , \; j = 3,4 \, ... \, N \; .
\label{adef}
\end{equation}
In its dual form the partition sum $Z$ is a sum over the dual variables $m_{x,\nu} \in \mathds{Z}$, $\overline{m}_{x,\nu} \in \mathds{N}_0$ 
and $n_{x,\nu}^{(j)} \in \mathds{N}_0, j = 3,4 \, ... \, N$. 
These are subject to constraints at each site $x$. The constraints are written explicitly in the first line
of (\ref{eq_dual_z}) using Kronecker deltas $\delta(n)$ and the evenness function $E(n)$:
\begin{equation}
\delta(n) \; \equiv \; \left\{
\begin{array}{cc} 
1 ~~~~ & n=0 \\
0 ~~~~ & \mbox{else}
\end{array} \right. 
\quad , \qquad 
E(n) \; \equiv \; \left\{
\begin{array}{cc}
1 ~~~~ &n~\mbox{even} \\
0 ~~~~ &n~\mbox{odd}
\end{array} \right. \; .
\end{equation}
The second line in (\ref{eq_dual_z}) gives the weight of a configuration $\{m,\overline{m},n^{(j)}\}$: The first factor contains the coupling to the
chemical potential, which now obviously is real and positive in the dual representation. The second factor in the weight comes from the expansion
of the exponential functions and the third factor is the contribution from the integral over the local spin variables. 

Note that by explicitly writing the constraints in the first line of (\ref{eq_dual_z}) the $a_x^{(j)}$ are 
forced to be even by the evenness functions $E(a_x^{(j)})$ and we can omit the
distinction of the two cases in (\ref{integrali}), using its first line only.
Inserting these integrals in the product in the last term of (\ref{eq_dual_z}) one finds that one term in the numerator cancels the denominator
of the previous factor, leading to the final form of the dual partition function 
\begin{eqnarray}
Z & = & \!\!\! \sum_{\{m,\overline{m},n^{(j)}\!\}}  \left(
\prod_x  \, \delta\Big(\sum_\nu[m_{x,\nu}-m_{x-\hat{\nu},\nu}]\Big) ~ \prod_{j=3}^N E\left(a_x^{(j)}\right) \right)
\label{eq_dual_zfinal} \\
& \times &\quad  
e^{-\mu\sum_x m_{x,d}} 
\left( \prod_{x,\nu} \frac{(J/2)^{|m_{x,\nu}|+2~\overline{m}_{x,\nu}}}{(|m_{x,\nu}|+\overline{m}_{x,\nu})!~\overline{m}_{x,\nu}!} 
\prod_{j=3}^{N} \frac{J^{\; n_{x,\nu}^{(j)}}}{n_{x,\nu}^{(j)}!}\right) 
\left( \prod_x \frac{\prod_{j=2}^N \Gamma\left( \frac{1}{2}\left(1+ \delta_{j,2} + a_x^{(j)} \right) \right)}
{\Gamma\left( \frac{1}{2}\left(N+\sum_{j=2}^{N} a_x^{(j)}\right) \right)}\right),
\nonumber 
\end{eqnarray}
where the $a^{(j)}_x$ are given in (\ref{adef}).

The final result (\ref{eq_dual_zfinal}) provides an exact reformulation of the partition function $Z$ in terms of the dual variables 
$m_{x,\nu} \in \mathds{Z}$, $\overline{m}_{x,\nu} \in \mathds{N}_0$ and $n_{x,\nu}^{(j)} \in \mathds{N}_0, j = 3,4 \, ... \, N$. All weight
factors are real and positive for arbitrary values of the chemical potential $\mu$ and Monte Carlo simulations are possible without
complex action problem. Large values of the dual variables are suppressed by the factorials coming from the expansion of the 
exponential function (and the $\Gamma$-factors do not change this).

It is important to note that the configurations of the dual variables are subject 
to constraints for the $m_{x,\nu}$ and the $n_{x,\nu}^{(j)}, j = 3,4\, ... \,N$. For the variables $m_{x,\nu}$ the Kronecker deltas
imply 
\begin{equation}
\label{divfree_on}
\sum_\nu[m_{x,\nu}-m_{x-\hat{\nu},\nu}] \; = \; 0 \; \; \forall x \; \; \;\;  \Longleftrightarrow \;\;\;\; \nabla \vec{m}_x \; = \; 0 \; \; \forall x \; .
\end{equation}
Obviously the lhs.\ of that equation is a discrete version of the divergence $\nabla \vec{m}_x$ and the constraint simply implies
that the discrete current $\vec{m}_x \in \mathds{Z}^d$ is conserved. Thus the admissible 
configurations of the variables $m_{x,\nu}$ (i.e., the configurations that obey the constraints) are closed loops of flux. 

This also gives rise to an interesting geometrical interpretation of the terms coupling to the chemical potential: In (\ref{eq_dual_zfinal}) the chemical
potential couples to $\sum_x m_{x,d}$, i.e., the total flux of $m_{x,\nu}$ in direction $\nu = d$ (i.e., the time direction with extension $N_t$). 
Since the $m$-flux is conserved the expression $\sum_x m_{x,d}$ corresponds to $N_t$ times the total winding number of the loops of $m$-flux
around the compactified time. Thus we find that the chemical potential $\mu$ couples to $\sum_x m_{x,d} \equiv N_t Q$, 
where the net particle number $Q$ is identified with the total winding number of the loops of $m$-flux around compactified time. 
In the bosonic case at hand this number is not restricted, but can take any integer value. Moreover, writing $m_{x,\nu}$ as the curl of another
integer-valued field, one can show that the chemical potential couples to the topological charge counting kinks in the dual Sine-Gordon model 
\cite{falktin}.
 
For the other set of constrained variables $n_{x,\nu}^{(j)} \in \mathds{N}_0, j = 3,4 \, ... \, N$, the constraints imply that the sums
$a_{x}^{(j)} = \sum_\nu [ n_{x,\nu}^{(j)}+n_{x-\hat{\nu},\nu}^{(j)} ]$ have to be even for all $x$ and $j = 3,4 \, ... \, N$. This can be written as
\begin{equation}
\sum_\nu[n^{(j)}_{x,\nu}+n^{(j)}_{x-\hat{\nu},\nu}] \;\;\; \mbox{even} \; \; \forall x \; \; \;\;  \Longleftrightarrow \;\;\;\; 
\sum_\nu[n^{(j)}_{x,\nu}-n^{(j)}_{x-\hat{\nu},\nu}] \;\;\; \mbox{even} \; \; \forall x \; \; \;\;  \Longleftrightarrow \;\;\;\; 
\nabla \vec{a}^{\,(j)}_x \;\;\; \mbox{even}  \; \; \forall x \; .
\end{equation}
In the second step we have changed the plus in the individual terms 
to a minus which corresponds to adding an even number on both sides. As a result
the constraints can again be written as a divergence and the constraint implies that the divergence $\nabla \vec{a}^{\,(j)}_x, \, 
j = 3,4 \, ... \, N$ has to vanish only modulo 2 (which 
corresponds to a $\mathds{Z}_2$ symmetry).  

\section{Dual representation of CP($N\!-\!1$) models with chemical potential}

Again the first step in the dualization is to write the spins $\vec{z}(x) \in \mathds{C}^N$ with $|\vec{z}(x)|^2 = \sum_{j=1}^N{|z_j(x)|^2} = 1$ 
using suitable coordinates. In particular we write the components $z_j(x)$ with polar coordinates in the complex plane,
\begin{equation}
z_j(x) \; = \; e^{i \varphi_j(x)} \, r_j(x) \quad \mbox{with} \quad \varphi_j(x) \in [0, 2\pi) \; , \; r_j(x) \in [0,1] 
\quad \mbox{and} \quad \sum_{j=1}^N{r_j(x)^2} \; = \; 1\; .
\label{polar}
\end{equation}
The moduli $r_j(x)$ are subject to the normalization $\sum_{j=1}^N{r_j(x)^2} \; = \; 1$ which restricts them to the $N-1$-sphere as in the 
O($N$) case. However, in addition we have the condition $r_j(x) \geq 0$ in order to obtain a unique representation with
the polar coordinates (\ref{polar}) chosen in the complex plane. We can parameterize the $r_j(x)$ again with generalized spherical coordinates,  
\begin{eqnarray}
r_1(x) &=& \sin\theta_N(x) \ldots \sin\theta_4(x) \sin\theta_3(x) \sin\theta_2(x)\; , \nonumber \\
r_2(x) &=& \sin\theta_N(x) \ldots \sin\theta_4(x) \sin\theta_3(x) \cos\theta_2(x)\; , \nonumber \\
r_3(x) &=& \sin\theta_N(x) \ldots \sin\theta_{{4}}(x) \cos\theta_3(x) \; , \nonumber \\
\vdots && \vdots \nonumber \\
r_{N-1}(x) &=& \sin\theta_N(x) \cos\theta_{N-1}(x) \; , \nonumber \\
r_N(x) &=& \cos\theta_N(x) \; ,
\end{eqnarray}
but now all angles $\theta_j(x), j = 2,4 \, ... \, N$ run only over the interval $[0,\pi/2]$, such that the $r_j(x)$ are 
non-negative. With this parameterization we can write the path integral measure $\mathcal{D}[\vec{z}\,]$ explicitly as
(we again omit parts of the normalization which drop out in expectation values anyway).
\begin{equation}
\label{cpn_measure}
\int \!\! \mathcal{D}[\vec{z}\,] =  \prod_x \!\! \left( \prod_{j=1}^N \int_0^{2\pi} \frac{d\varphi_j(x)}{2\pi} \! \right) \;
{\prod_{k=2}^N \,  \int_0^{\pi/2} d \theta_k(x) \; \cos{\theta_k(x)} \, \sin(\theta_k(x))^{2k - 3}}  .
\end{equation}
For the CP($N\!-\!1$) case we decompose the action (\ref{eq_actioncpn}) in the form
\begin{eqnarray}
&& S[\vec{z},U] \;  =  \; - \sum_{j=1}^N \sum_x \sum_{\nu = 1}^d \Big[ S_\nu^{\,(j)}(x) \, + \, \overline{S}_\nu^{\,(j)}(x) \Big] \; , 
 \\
&& S_\nu^{\,(j)}(x) \; = \; J \, e^{- \mu_j \, \delta_{\nu,d}} z_j(x)^* \, U_\nu(x) z_j(x+\hat{\nu}) \; \; , \; \; 
\overline{S}_\nu^{\,(j)}(x) \; = \; J \, e^{\, \mu_j \, \delta_{\nu,d}} z_j(x) \, U_\nu(x)^* z_j(x+\hat{\nu})^* \; .
\nonumber
\end{eqnarray}
The link variables $U_\nu(x)$ are for now considered as fixed external fields (we integrate them out later). For finding the partition sum 
$Z[U]$ in a background configuration of the link variables we again factorize the Boltzmann weight and expand each term individually:  
\begin{eqnarray}
Z[U] & \!\!\! = \!\!\! & 
\int \!\! \mathcal{D}[\vec{z}\,] \, e^{-S[\vec{z},U]} =  
\int \!\! \mathcal{D}[\vec{z}\,]  \! \prod_{x,\nu,j} e^{S^{\,(j)}_\nu(x)} \, e^{\overline{S}^{\,(j)}_\nu(x)} = 
\int \!\! \mathcal{D}[\vec{z}\,] \! \prod_{x,\nu,j}  \; \sum_{k^{(j)}_{x,\nu}=0}^\infty \!\!\!\!
\frac{\Big(\!S^{(j)}_\nu(x)\!\Big)^{k^{(j)}_{x,\nu}}}{k^{(j)}_{x,\nu}!} \!\!
\sum_{\overline{k}^{(j)}_{x,\nu}=0}^\infty \!\!\!\!
\frac{\Big(\overline{S}^{(j)}_\nu(x)\!\Big)^{\overline{k}^{(j)}_{x,\nu}}}{\overline{k}^{(j)}_{x,\nu}!} 
\nonumber \\
& \!\!\! = \!\!\! & \!\!\!\!\!\! \sum_{\{ k^{(j)}, \overline{k}^{(j)} \} } 
\left( \prod_{x,\nu,j} \frac{ J^{k_{x,\nu}^{(j)} } }{ k_{x,\nu}^{(j)} ! } \frac{ J^{\overline{k}_{x,\nu}^{(j)} } }{ \overline{k}_{x,\nu}^{(j)} ! } \right) 
\left( \prod_{x,\nu} U_\nu(x)^{ \sum_j [ k_{x,\nu}^{(j)} - \overline{k}_{x,\nu}^{(j)} ] } \right) \,
e^{-\sum_{x,j} \mu_j \, [ k_{x,d}^{(j)} - \overline{k}_{x,d}^{(j)} ] } \; \mathcal{I} [\{ k^{(j)}, \overline{k}^{(j)} \}] \, .
\label{expansioncp1}
\end{eqnarray}
By $\sum_{\{ k^{(j)}, \overline{k}^{(j)} \} }$ we denote the sum over all configurations of the expansion indices
$k_{x,\nu}^{(j)}, \overline{k}_{x,\nu}^{(j)} \in \mathds{N}_0$ assigned to the links of the lattice. The remaining 
integral over the powers of spin components at each site of the lattice can be factorized into 
two parts, $\mathcal{I} [\{ k^{(j)}, \overline{k}^{(j)} \}] = \mathcal{I}_{\varphi} [\{ k^{(j)}, \overline{k}^{(j)} \}] \;
\mathcal{I}_{r} [\{ k^{(j)}, \overline{k}^{(j)} \}]$, which correspond to the integration over the phases and the 
moduli of the spin components. 

The integral over the phases $\varphi_j(x)$ is rather trivial:
\begin{eqnarray}
\mathcal{I}_{\varphi} [\{ k^{(j)}\!, \overline{k}^{(j)} \}]  & = &  
\prod_{x,j}  \int_0^{2\pi} \!\! \frac{d \varphi_j(x)}{2\pi} e^{i \varphi_j(x) \sum_\nu \big[ \overline{k}_{x,\nu}^{(j)} \!-  k_{x,\nu}^{(j)}
- ( \overline{k}_{x-\hat{\nu},\nu}^{(j)}\! -  k_{x-\hat{\nu},\nu}^{(j)} ) \big] } 
\nonumber \\
& = &  
\prod_{x,j}  \delta \left( \! \sum_\nu \Big[  {k}_{x,\nu}^{(j)}\! - \! \overline{k}_{x,\nu}^{(j)}
- ( {k}_{x-\hat{\nu},\nu}^{(j)}\! - \! \overline{k}_{x-\hat{\nu},\nu}^{(j)} ) \Big]\! \right) ,
\end{eqnarray}
it simply gives rise to a product of Kronecker deltas for all sites. At every site the Kronecker deltas enforce the constraint
\begin{equation}
\sum_\nu \Big[ {k}_{x,\nu}^{(j)}\! - \! \overline{k}_{x,\nu}^{(j)}
- ( {k}_{x-\hat{\nu},\nu}^{(j)}\! - \! \overline{k}_{x-\hat{\nu},\nu}^{(j)} ) \Big] = 0 \quad , \quad j = 1,2 \, ... \, N \; .
\label{constraint1cp}
\end{equation} 
The integral over the parameterized moduli $r_j(x)$ is given by
\begin{equation}
\label{constraint2cp}
\mathcal{I}_r  [\{ k^{(j)}\!, \overline{k}^{(j)} \}]  = \prod_x 
{\prod_{k=2}^N
\int_0^{\pi/2}  d\theta_k(x)  \cos\theta_k(x)^{1 + a^{(k)}_x} \sin\theta_k(x)^{2k -3 + \sum_{j=1}^{k-1} a^{(j)}_x}}
 \end{equation}
where we use the abbreviation
\begin{equation}
 a_{x}^{(j)} \; = \; \sum_\nu \Big[ k_{x,\nu}^{(j)} + \overline{k}_{x,\nu}^{(j)}  + k_{x-\hat{\nu},\nu}^{(j)} + \overline{k}_{x-\hat{\nu},\nu}^{(j)} \Big] 
 \quad , \quad j = 1,2 \, ... \, N \; .
\label{adefcp}
\end{equation}
All the integrals that appear in (\ref{constraint2cp}) are related to (\ref{integrali}) but the integration runs only up to $\pi/2$. 
Thus the result is always given by the first line of (\ref{integrali}) with an extra factor of $1/2$.
When taking the product of all the integrals that build up  (\ref{constraint2cp}),
again the numerator of a term and one of the denominators of the subsequent term cancel. The integral 
$\mathcal{I}_r  [\{ k^{(j)}\!, \overline{k}^{(j)} \}]$ assumes the simple form (we dropped an irrelevant overall factor $2^{-NV}$)
\begin{equation}
\mathcal{I}_r  [\{ k^{(j)}\!, \overline{k}^{(j)} \}]  \; = \; \prod_x \frac{\prod_{j=1}^N \Gamma\left( \frac{1}{2}\left({2} + a_x^{(j)} \right) \right)}
{\Gamma\left( \frac{1}{2}\left({2N}+\sum_{j=1}^{N} a_x^{(j)}\right) \right)} \; .
\end{equation}
Note that here the integrals over the $\theta_j(x)$ did not generate any additional constraints -- the constraints  (\ref{constraint1cp})
from integrating the $\varphi_j(x)$ already imply the evenness of the $a_{x}^{(j)}$.

As before we can simplify the expressions further by rewriting the dual variables 
$k^{(j)}_{x,\nu}, \overline{k}^{(j)}_{x,\nu} \in \mathds{N}_0$  in terms of 
new dual variables $m^{(j)}_{x,\nu} \in \mathds{Z}$ and $\overline{m}^{(j)}_{x,\nu} \in \mathds{N}_0$, 
\begin{equation}
k_{x,\nu}^{(j)}-\overline{k}^{(j)}_{x,\nu} \equiv m^{(j)}_{x,\nu}  \;\;, \;\;\; m^{(j)}_{x,\nu} \in \mathds{Z} \quad \mbox{and} \quad
k_{x,\nu}^{(j)}+\overline{k}^{(j)}_{x,\nu} \equiv |m^{(j)}_{x,\nu}|+2~\overline{m}^{(j)}_{x,\nu} \;\; , \;\;\;  \overline{m}^{(j)}_{x,\nu} \in \mathds{N}_0 \; .
\end{equation}
The sums $a_x^{(j)}$ defined in (\ref{adefcp}) turn into
\begin{equation}
 a_{x}^{(j)} \; = \; \sum_\nu \Big[ |m_{x,\nu}^{(j)}|+ 2\overline{m}_{x,\nu}^{(j)}  + |m_{x-\hat{\nu},\nu}^{(j)}| + 2\overline{m}_{x-\hat{\nu},\nu}^{(j)} \Big] 
 \quad , \quad j = 1,2 \, ... \, N \; ,
\label{adefcp2}
\end{equation}
and we find for the partition sum $Z[U]$ in the background field of the $U_\nu(x)$:
\begin{eqnarray}
Z[U] & = & \sum_{\{ m^{(j)}, \overline{m}^{(j)} \} } 
\left( \prod_{x,j} \! \delta \!\left( \sum_\nu [  {m}_{x,\nu}^{(j)} - {m}_{x-\hat{\nu},\nu}^{(j)} ] \right) \right)
\left( \prod_{x,\nu,j} 
\frac{ J^{ | m_{x,\nu}^{(j)} |  +  2 \overline{m}_{x,\nu}^{(j)} } } 
{ ( | m_{x,\nu}^{(j)} | + \overline{m}_{x,\nu}^{(j)} )! \; \overline{m}_{x,\nu}^{(j)} ! } \right)  
\left( \prod_{x,j} e^{ - \mu_j  m _{x,d}^{(j)} } \right) 
\nonumber \\
&& \times 
\left( \prod_x \frac{\prod_{j=1}^N \Gamma\left( \frac{1}{2}\left({2} + a_x^{(j)} \right) \right)}
{\Gamma\left( \frac{1}{2}\left({2N}+\sum_{j=1}^{N} a_x^{(j)}\right) \right)} \right)
\left( \prod_{x,\nu} U_\nu(x)^{ \sum_j m_{x,\nu}^{(j)} } \right) .
\end{eqnarray}
The final step\footnote{Actually this could have been done at any stage of the derivation.} is  to integrate the link variables $U_\nu(x) = e^{i A_\nu(x)}$ with the measure $\int \! \mathcal{D}[U] \! = \prod_{x,\nu} \int_0^{2\pi}
\! \frac{d A_\nu(x) }{2 \pi}\!$, which gives rise to another set of constraints:
\begin{equation}
\label{link_integration}
\int \mathcal{D}[U] \, \prod_{x,\nu} U_\nu(x)^{ \sum_j m_{x,\nu}^{(j)} } \; = \; 
\prod_{x,\nu} \int_0^{2\pi} \! \frac{d A_\nu(x) }{2 \pi} \, U_\nu(x)^{ \sum_j m_{x,\nu}^{(j)} }
\; = \; \prod_{x,\nu} \delta \left( \sum_j m_{x,\nu}^{(j)} \right) \; .
\end{equation}
The final result for the CP($N\!-\!1$) partition sum in the dual representation thus reads   
\begin{eqnarray}
Z & = & \sum_{\{ m^{(j)}, \overline{m}^{(j)} \} } 
\left( \prod_{x,j} \! \delta \!\left( \sum_\nu [  {m}_{x,\nu}^{(j)} - {m}_{x-\hat{\nu},\nu}^{(j)} ] \right) \right)
\left( \prod_{x,\nu} \delta \left( \sum_j m_{x,\nu}^{(j)} \right) \right) 
\label{cpnzfinal} \\
& \times &
\left( \prod_{x} e^{ \, - \sum_{j=1}^N \mu_j  m _{x,d}^{(j)} } \right)
\left( \prod_{x,\nu,j} 
\frac{ J^{ | m_{x,\nu}^{(j)} |  +  2 \overline{m}_{x,\nu}^{(j)} } } 
{ ( | m_{x,\nu}^{(j)} | + \overline{m}_{x,\nu}^{(j)} )! \; \overline{m}_{x,\nu}^{(j)} ! } \right)   
\left( \prod_x \frac{\prod_{j=1}^N \Gamma\left( \frac{1}{2}\left({2} + a_x^{(j)} \right) \right)}
{\Gamma\left( \frac{1}{2}\left({2N}+\sum_{j=1}^{N} a_x^{(j)}\right) \right)} \right) ,
\nonumber
\end{eqnarray}
with the $a_x^{(j)}$ given by (\ref{adefcp2}). Only the dual variables $m_{x,\nu}^{(j)} \in \mathds{Z}$ are subject to constraints,
while the $\overline{m}_{x,\nu}^{(j)} \in \mathds{N}_0$ are unconstrained. All constraints are collected in the first line of
(\ref{cpnzfinal}): The first factor implies that all discrete currents $\vec{m}_x^{(j)}$ are divergence-free, i.e., 
$\nabla \vec{m}_x^{(j)} = \sum_\nu [  {m}_{x,\nu}^{(j)} - {m}_{x-\hat{\nu},\nu}^{(j)} ] = 0 \; \forall x, j$, 
similar to the O(N) case in Eq.~(\ref{divfree_on}). We stress that the condition 
of vanishing divergence holds individually for all $j = 1,2 \, ... \, N$. A second constraint ties together the components $j = 1,2 \, ... \, N$: 
At each link the sum $\sum_j m_{x,\nu}^{(j)}$ of the dual variables $m_{x,\nu}^{(j)} \in \mathds{Z}$ has to vanish. Note that this implies that 
for the case of identical chemical potentials $\mu_1 = \mu_2 = \, ... \, =\mu_N \equiv \mu$ the model is independent of $\mu$. This is just reflecting 
the fact that the chemical potential for the U(1) gauge symmetry cannot affect the theory. The reason for this is that the gauge field is non-dynamical,
 and the Gauss law is simply a statement that the charge density of this U(1) symmetry vanishes at every point, 
 and the total U(1) charge must be zero for all physical states of the model. 

The weight factors are collected in the second line of (\ref{cpnzfinal}). It is obvious that all of them are real and positive also for arbitrary values 
of the chemical potential and the complex action problem is solved in the dual form. As for the O($N$) case, the chemical potentials couple
to the time component of the $m^{(j)}$ currents and combined with the flux conservation (i.e., the condition $\nabla \vec{m}_x^{(j)} = 0$)
this implies that the charge $Q_j$ the chemical potential $\mu_j$ couples to is given by the winding number of the $m^{(j)}$-flux around
the compact time direction. 

We remark that in two dimensions ($d\!=\!2$) one may consider a topological term by including an additional Boltzmann factor 
$e^{i \theta Q[U]}$ with $Q[U] = (2\pi)^{-1}\! \sum_x  \mbox{Im} \, U_1(x) U_2(x\!+\!\hat1) U_1(x\!+\!\hat2)^* U_2(x)^*$. This introduces an
additional source for an imaginary part in the action. It turns out, that this latter complex action problem cannot be removed completely in our 
dualization (the same conclusion was found also in the alternative dualization in \cite{wolff_CPN}). This is an interesting finding, since
for U(1) lattice field theories that contain the same topological term but also an action term for the gauge fields, the dualization does indeed 
solve the complex action problem from the vacuum term \cite{theta1,theta2}.

\section{Remark on the relation of the dual O(3) and CP(1) models}

An interesting question is how the dual representations of the O(3) and the CP(1) models are related. These are the lowest-$N$ cases of the two 
models and in the conventional continuum representation can be transformed into each other (in this discussion we restrict 
ourselves to vanishing chemical potential). In the continuum one can show via $r_I=z_i^*(\sigma_I)_{ij}z_{j}$ with $\sigma$ the Pauli matrices, 
that the O(3) action $(\partial_\nu\vec{r})^2/2$ turns into twice the quartic CP(1) action 
$(\partial_\nu z)^\dagger(\partial_\nu z)+(z^\dagger\partial_\nu z)^2$. The latter occurs from the action with gauge fields 
$(D_\nu z)^\dagger(D_\nu z)$, $D_\nu=\partial_\nu+iA_\nu$, after integrating out the gauge field.

On the lattice one can do the same. Under the field mapping above, the lattice O(3) actions (\ref{action0}), (\ref{eq_action_mu}) turn 
into actions quartic in the CP(1) fields closely resembling the continuum quartic CP(1) action. On the other hand, the 
Boltzmann factor with the CP(1) action 
(\ref{eq_actioncpn}) (set $N=2$), after integrating out the link field can be shown to be a product of Bessel functions and not of the form 
$\exp(-S)$ with $S$ a polynomial action in the CP(1) fields. Hence, the O(3) actions from which the dualizations start are different and 
the same is expected for the dualized partition functions. The differences should of course vanish in the continuum limit. Note that both dualizations 
possess three dual variables, $(m,\,\overline{m},\,n^{(3)})$ in the O($3$) case vs. 
$(m^{(1)},\,\overline{m}^{(1)}, \, \overline{m}^{(2)} )$ in the CP($1$) 
case (having solved the second constraint by putting $m^{(2)}=-m^{(1)})$. These two dualizations for the O(3) model could have different 
computational features and might highlight different aspects of our understanding of dualizations in general.

\section{Summary and comments}

In this paper we have derived dual representations for the O($N$) and CP($N\!-\!1$) models with chemical potentials. The dual representations
use $N$ sets (respectively $2N$ sets for the CP($N\!-\!1$) case) of integer valued dual link variables. Some of the dual link variables obey
constraints, in particular they have vanishing divergence, i.e., conserved flux. The chemical potential couples to charges which in the dual 
picture can be interpreted as the winding numbers of the corresponding conserved flux around the compact time direction. All weight factors 
are real and positive and in the dual representation the complex action problem is solved. Suitable dual Monte Carlo strategies were 
discussed in the literature (see, e.g., \cite{review1,review2,worm,surfaceworm}, \cite{endres,wolff_ON,wolff_CPN,shailesh_O2} for examples) and for the case discussed here we are currently preparing
a paper \cite{wip} presenting the results of a dual simulation in the representation derived here. 

Vacuum expectation values of observables can be obtained easily in the dual representation: Derivatives of $\ln Z$ 
with respect to the couplings $J$ and $\mu_j$ give rise to bulk observables and their moments which can be used for determining 
the various transitions in the models \cite{wip}. 
In addition one can go through the same dualization steps also for chemical potentials $\mu_j(x)$ that depend on the space-time arguments $x$. 
These then may be used as source terms and after taking suitable derivatives of $\ln Z$ one obtains the corresponding $n$-point functions. 

In the introduction we have discussed that there is no general strategy for finding a dual representation. Studying various
dual forms of different models thus provides new tools for attacking more challenging non-abelian symmetries, in particular gauge theories
with non-abelian gauge groups. However, not only finding more tools for dualization is important, but also understanding the dual variables  
is an interesting issue per se. Here we have demonstrated that different dual representations exist for the same model, 
(in detail we discussed the case O(3)$\,\cong\,$CP(1)), and one thus might try to 
analyze what constitutes the key ingredients of a dual representation. Understanding these key characteristics 
might open the possibility of formulating 
lattice field theories directly in the dual framework of matter fluxes and surfaces for gauge fields. Clearly this is a rather speculative
perspective at the moment, but it might become more relevant with every successful dualization of a lattice field theory. 

\vskip5mm
\noindent
{\bf Acknowledgments:} 
FB is supported by the DFG (BR 2872/6-1) and TK by the Austrian Science Fund, 
FWF, DK {\sl Hadrons in Vacuum, Nuclei, and Stars} (FWF DK W1203-N16).  
Furthermore this work is partly supported by the Austrian Science Fund FWF Grant.\ Nr.\ I 1452-N27 and by the 
DFG TR55, {\sl ''Hadron Properties from Lattice QCD''}.  


\begin{thebibliography}{12}

\bibitem{review1}
  S.~Chandrasekharan,
  PoS LATTICE {\bf 2008} (2008) 003
  [arXiv:0810.2419 [hep-lat]].

\bibitem{review2}
  C.~Gattringer,
  PoS LATTICE {\bf 2013} (2014) 002
  [arXiv:1401.7788 [hep-lat]].
  
\bibitem{worm}
  N.~Prokof'ev and B.~Svistunov,
  Phys.\ Rev.\ Lett.\  {\bf 87} (2001) 160601.

\bibitem{surfaceworm}
  Y.D.~Mercado, C.~Gattringer and A.~Schmidt,
  Comput.\ Phys.\ Commun.\  {\bf 184} (2013) 1535
  [arXiv:1211.3436 [hep-lat]].
  
  \bibitem{Eto:2004rz} 
  M.~Eto, Y.~Isozumi, M.~Nitta, K.~Ohashi and N.~Sakai,
  Phys.\ Rev.\ D {\bf 72}, 025011 (2005)
  [hep-th/0412048].
 
 \bibitem{Eto:2006mz} 
  M.~Eto, T.~Fujimori, Y.~Isozumi, M.~Nitta, K.~Ohashi, K.~Ohta and N.~Sakai,
  Phys.\ Rev.\ D {\bf 73}, 085008 (2006)
  [hep-th/0601181].
  
\bibitem{Bruckmann:2007zh}
  F.~Bruckmann,
  Phys.\ Rev.\ Lett.\  {\bf 100} (2008) 051602
  [arXiv:0707.0775 [hep-th]].
 
 \bibitem{Brendel:2009mp} 
  W.~Brendel, F.~Bruckmann, L.~Janssen, A.~Wipf and C.~Wozar,
  Phys.\ Lett.\ B {\bf 676}, 116 (2009)
  [arXiv:0902.2328 [hep-th]].
  
  \bibitem{Harland:2009mf} 
  D.~Harland,
  J.\ Math.\ Phys.\  {\bf 50}, 122902 (2009)
  [arXiv:0902.2303 [hep-th]].

\bibitem{Nitta:2014vpa}
  M.~Nitta,
  JHEP {\bf 1503} (2015) 108
  [arXiv:1412.7681 [hep-th]].

\bibitem{Nitta:2015tua}
  M.~Nitta,
  arXiv:1503.06336 [hep-th].

\bibitem{Dunne:2012ae} 
  G.V.~Dunne and M.~Unsal,
  JHEP {\bf 1211}, 170 (2012)
  [arXiv:1210.2423 [hep-th]].
  
\bibitem{Dunne:2012zk} 
  G.V.~Dunne and M.~Unsal,
  Phys.\ Rev.\ D {\bf 87}, 025015 (2013)
  [arXiv:1210.3646 [hep-th]].
 
\bibitem{Cherman:2013yfa}
  A.~Cherman, D.~Dorigoni, G.V.~Dunne and M.~Unsal,
  Phys.\ Rev.\ Lett.\  {\bf 112} (2014) 021601
  [arXiv:1308.0127 [hep-th]].

\bibitem{Dunne:2015ywa} 
  G.V.~Dunne and M.~Unsal,
  arXiv:1505.07803 [hep-th].

\bibitem{endres}
 M.G.~Endres,
  Phys.\ Rev.\ D {\bf 75} (2007) 065012
  [hep-lat/0610029];
%
  PoS LAT06 (2006) 133
  [hep-lat/0609037].

\bibitem{wolff_ON}
  U.~Wolff,
  Nucl.\ Phys.\ B {\bf 824} (2010) 254
  [E: Nucl.\ Phys.\  {\bf 834} (2010) 395]
  [arXiv:0908.0284 [hep-lat]].

\bibitem{wolff_CPN}
  U.~Wolff,
  Nucl.\ Phys.\ B {\bf 832} (2010) 520
  [arXiv:1001.2231 [hep-lat]].
  
\bibitem{shailesh_O2}
  D.~Banerjee and S.~Chandrasekharan,
  Phys.\ Rev.\ D {\bf 81} (2010) 125007
  [arXiv:1001.3648 [hep-lat]].

\bibitem{alles}  
  C.~Torrero, O.~Borisenko, V.~Kushnir, B.~Alles and A.~Papa,
  PoS LATTICE {\bf 2013} (2014) 338
  [arXiv:1311.4292 [hep-lat]].
   
\bibitem{falktin}
  F.~Bruckmann, T.~Sulejmanpasic,
  Phys.\ Rev.\ D {\bf 90} (2014) 10,  105010
  [arXiv:1408.2229 [hep-th]].

\bibitem{meurice}  
L.~P.~Yang, Y.~Liu, H.~Zou, Z.~Y.~Xie and Y.~Meurice,
  arXiv:1507.01471 [cond-mat.stat-mech].
  
\bibitem{dualsun_1}
  N.~D.~Hari Dass,
  Nucl.\ Phys.\ Proc.\ Suppl.\  {\bf 83} (2000) 950
  [hep-lat/9908049].

\bibitem{dualsun_2}
   N.~D.~Hari Dass,
  Nucl.\ Phys.\ Proc.\ Suppl.\  {\bf 94} (2001) 665
  [hep-lat/0011047].

\bibitem{dualsun_3}
   N.~D.~Hari Dass and D.~S.~Shin,
  Nucl.\ Phys.\ Proc.\ Suppl.\  {\bf 94} (2001) 670
  [hep-lat/0011038].

\bibitem{dualsun_4}
  J.~W.~Cherrington, D.~Christensen and I.~Khavkine,
  Phys.\ Rev.\ D {\bf 76} (2007) 094503
  [arXiv:0705.2629 [hep-lat]].
 
\bibitem{dualsun_5}
 J.~W.~Cherrington,
  Nucl.\ Phys.\ B {\bf 794} (2008) 195
  [arXiv:0710.0323 [hep-lat]].
 
\bibitem{dualsun_6}
 J.~W.~Cherrington,
  PoS LATTICE {\bf 2008} (2008) 050
  [arXiv:0810.0546 [hep-lat]].
  
\bibitem{dualsun_7}
  J.~W.~Cherrington,
  Nucl.\ Phys.\ B {\bf 835} (2010) 29
  [arXiv:0908.1889 [hep-lat]].
   
\bibitem{dualsun_8}
  J.~W.~Cherrington,
  Nucl.\ Phys.\ B {\bf 835} (2010) 51
  [arXiv:0908.1893 [hep-lat]].

\bibitem{Hasenfratz:1990ab} 
  P.~Hasenfratz and F.~Niedermayer,
  Phys.\ Lett.\ B {\bf 245}, 529 (1990).

\bibitem{cpnlat}
  P.~Di Vecchia, A.~Holtkamp, R.~Musto, F.~Nicodemi, R.~Pettorino,
  Nucl.\ Phys.\ B {\bf 190} (1981) 719.

\bibitem{zoller}  
  C.~Laflamme, W.~Evans, M.~Dalmonte, U.~Gerber, H.~Mej'a-D'az, W.~Bietenholz, U.-J.~Wiese and P.~Zoller,
  arXiv:1507.06788 [quant-ph].
 
\bibitem{theta1}
  C.~Gattringer, T.~Kloiber, V.~Sazonov,
  Nucl.\ Phys.\ B {\bf 897} (2015) 732
  [arXiv:1502.05479 [hep-lat]].

\bibitem{theta2}
  T.~Kloiber and C.~Gattringer,
  PoS Lattice 2014 (2015) 345 [arXiv:1410.3216 [hep-lat]].

\bibitem{wip}
  F.~Bruckmann, C.~Gattringer, T.~Kloiber and T.~Sulejmanpasic,
  work in preparation.
 
\end{thebibliography}
\end{document}